# Room temperature self-assembly of cation-free guanine quartet network nucleated from Mo-induced defect on decorated Au(111) with graphene nanoribbons

Amirreza Ghassami,[†] Elham Oleiki,[†] Dong Yeon Kim,[†] Hyung-Joon Shin,*[‡] Geunsik Lee*[†] and Kwang S. Kim*[†]

[†] Center for Superfunctional Materials, Department of Chemistry, Ulsan National Institute of Science and Technology (UNIST), 50 UNIST-gil, Ulsan 44919, Republic of Korea

[‡] Department of Materials Science and Engineering, Ulsan National Institute of Science and Technology (UNIST), 50 UNIST-gil, Ulsan 44919, Republic of Korea

Email: gslee@unist.ac.kr; shinhj@unist.ac.kr; kimks@unist.ac.kr

Guanine-quadruplex, consisting of several stacked guanine-quartets (GQs), has emerged as an important category of novel molecular targets with applications from nanoelectronic devices to anticancer drugs. Incorporation of metal cations into GQ structure is utilized to form stable G-quadruplexes, while no other passage has been reported yet. Here we report the room temperature (RT) molecular self-assembly of extensive metal-free GQ networks on Au(111) surface. Surface defect induced by an implanted molybdenum atom within Au(111) surface is used to nucleate and stabilize the cation-free GQ network. Additionally, the decorated Au(111) surface with 7-armchair graphene nanoribbons (7-AGNRs) results in more extensive GQ networks by curing the disordered phase nucleated from Au step edges spatially and chemically. Scanning tunneling microscopy/spectroscopy (STM/STS) and density functional theory (DFT) calculations confirm GQ networks' formation and unravel the nucleation and growth mechanism. This method stimulates cation-free G-quartet network formation at RT and can lead to stabilizing new emerging molecular self-assembly.

## Introduction

Self-assembly of organic molecules with sufficient stability is highly desirable to develop intriguingly organized materials or molecular architectures for diverse applications from nanoelectronic circuits to drug design.[1–4] The emergence of various RNA/DNA quadruplex structures is a vital discovery in biomolecular science for their antiproliferative behavior and gene-regulating features of non-canonical RNA/DNA secondary structures.[5–7] Among various RNA/DNA quadruplexes, guanine is the only purine generated in human cells to date[8,9], where stabilizing G-quadruplex has become an efficient way of designing small-molecule drugs.[10,11] When molecules are assembled on substrates, various forms of quartet (Q) networks have been observed for guanine (G), xanthine (X), and hypoxanthine (HX).[12–14] In view of molecular electronics, DNA G-quadruplex is a legitimate candidate for miniaturizing memory storages and electronic devices since it is capable of transporting electrical current and inherits DNA's versatile and programmable structure, while native double-stranded DNA produces no detectable signal.[1]

The assembly at surfaces is a suitable platform to study systematically how inter-molecular interaction and molecule-surface interaction interplay with each other to give rise to the resulting self-assembly network.[15] Several studies have been reported for nucleobases self-assembly on Au(111).[16–20] Mainly, the chain-like networks were observed primarily due to relatively strong intermolecular hydrogen bonds. In addition, the impact of impurity atoms on the network patterns has been studied. For example, the presence of metal atoms is not necessitated for the fabrication of X-quartets[14] but is required to form GQ and HX-Q networks.[13,21,22] A pristine form of GQ is electrostatically unfavourable as four oxygen atoms are located towards the central cage. Thus it needs to be stabilized by the presence of a caged cation,[22] either $K^+$ or $Na^+$ cations;[19,22,23] or otherwise a new GQ-M complex (or supramolecular structure) is produced after introducing transition metal adatoms to guanine molecular source.[14,21] In this regard, the cooperative effect of hydrogen bonds, responsible for Watson-Crick base pairing, is also unable to stabilize the metal-free GQ network on the pristine Au(111) surface in addition to resonance-assisted hydrogen bonding (RAHB) effect, even though considered presumably,[12] but revisited and corrected afterward.[13,14,21,24] Also, the geometry of a stabilized initial seed is an additional crucial factor in forming and stabilizing the molecular network, where the nucleation can proceed from either a flat terrace or edge region. To block surface diffusion towards undesirable nucleation sites, a heterogeneous admixture is employed.[25] Stabilizing pure GQ networks without incorporating metal cations boost opportunities to fabricate more diverse, innovative structures with novel applications in genetic modification, protein folding, and nanoelectronic circuits. Here we report RT self-assembly of cation-free GQ network on Au(111) by substituting minute surface Au atoms by Mo atoms, which shows robust thermal stability against annealing up to 500 K. The underlying mechanism is demonstrated to be the covalent interaction between G/9H and surface Mo from our corroborative confirmation with STS measurement and DFT calculations. Furthermore, enlarged domain size of GQ networks was observed by the bottom-up synthesis of GNRs prior to G growth. Our work provides a facile approach for enhancing the stability and domain size of GQ network, which might be applicable to two dimensional self-assembly of other nucleobases which have shown only limited stability and size.

## Results and discussion

### Guanine self-assembly on pristine Au(111)

Guanine has several molecular tautomers, and the two most stable ones are the canonical G/9H and the non-canonical G/7H forms.[22,26,27] These two forms, like many other tautomeric structures of organic compounds, transform to each other through the migration of an H atom and coexist when reaching a certain temperature or contacting with specific materials.[27–29] Previous works on the 2D assembly show that a GQ network is based on the canonical structure (G/9H) in the presence of alkali metals.[12,22] The guanine zigzag structure on Au(111), which is the thermodynamically most stable structure on surface, has double and triple intermolecular hydrogen bonds and is derived from G/7H tautomeric form with a mixture of left and right chirality.[12,16,17,22] In our experiments, after the deposition of guanine/9H molecules on pristine Au(111) at RT, the STM images show the disordered arrangement on surface (Fig. 1a). Annealing disordered guanine molecules on Au(111) at 400 K stimulates molecules to move on surface more freely and transform into the G/7H tautomeric structure, for which the STM images show a close-packed lattice of G/7H molecules as shown in Fig. 1a. According to our DFT calculations, individual guanine molecules adsorb weakly in a face-down geometry via van der Waals interaction. The vertical distance between the O atom and Au surface layer is 2.6 Å for G/9H and 3.2 Å for G/7H molecule, with the adsorption energies of -1.38 eV and -1.24 eV, respectively (Fig. 1b). This difference in adsorption energy of guanine tautomers can be evidenced by the fact that the G/9H has a much stronger molecular dipole moment than any other guanine tautomers, while the G/7H has the smallest molecular dipole moment among them.[30,31] Annealing helps adsorbed molecules find each other to form the most stable network connected by hydrogen bonding. The DFT predicted free-standing lattice formation energy is -1.75 eV/molecule for G/7H, stronger than -1.27 eV/molecule for G/9H (Fig. S1a-b). The difference (0.48 eV/molecule) is sufficient to stabilize the G/7H lattice over the G/9H lattice, where a single G/9H molecular adsorption was more stable than G/7H by 0.14 eV/molecule, which is in good agreement with the observed close-packed lattice after annealing at 400 K (Fig. 1a).

### Guanine self-assembly on Mo-doped Au(111)

Although the chemically stable Au(111) provides a suitable environment for the formation of molecular surface assemblies,[32,33] the formation of metal-free self-assembled GQ networks was assumed improbable on Au(111) surface.[13,21] Intraquartet hydrogen bonding alone is not sufficient for stabilizing a GQ structure, and the RAHB effect described earlier[12] was finally discarded.[24] Furthermore, even with metal ions on Au(111) surface, fabrication of an extensive GQ network is also not possible at RT. It needs annealing at a specific temperature;[13,14,21,22] therefore, a mixture of adsorbates (metal ions plus guanine molecules) find the most stable network structure.[34,35] When the molecule-surface interaction is weak, like the van der Waals interactions between guanine molecules and Au(111) surface, G/9H molecules have significant mobility unless ample kinetic energy is diminished by collisions at RT.[32] Consequently, molecules cannot selectively form an ordered hydrogen bonding network, but a random molecular structure would be formed (Fig. 1a).

For one-step fabrication of GQ network at RT, we stabilized the initial seed on terrace regions by doping Au(111) surface with a small amount of Mo atoms (Fig. 2a-c), where the atomically flat surface and Au(111) herringbone structure are maintained after depositing Mo clusters and annealing. In order to remove any trace of Mo adatoms that might remain on Au surface after deposition, the sample was annealed at 700 K for 30 min (Fig. S2-S3). Our DFT calculations show that the G/9H molecule makes a covalent bond with an embedded Mo atom at Au surface layer in a face-down geometry with -2.42 eV adsorption energy and a vertical distance of 2.0 Å between O and Mo atoms (Fig. 2d). As we calculated the formation energy of the GQ network in the presence of one Mo-O covalent bond per four G/9H molecules (Fig. 2e), the molecular network becomes more stable with the formation energy of -1.96 eV/molecule on Mo-doped Au surface, compared to the GQ network on the pristine Au(111) with the formation energy of -1.82 eV/molecule (Table 1). Also, when two or three Mo atoms are embedded in the surface layer, the formation of GQ network is slightly more enhanced as compared to the case of a single Mo (Fig. S4 and Table 1). However, this di-vacancy replacement would hardly take place as compared with the mono-vacancy replacement (Fig. S5-S6), and the mono-vacancy replacement is entropically favoured at high-temperature annealing (700 K). Overall, there will be a more probability of finding the GQ form's initial seeds on the Mo-doped Au(111) surface than the pristine Au(111) (Fig. 2a and S1c). As a result of intraquartet hydrogen bonding and covalent bonding between Mo at surface and O of G/9H in addition to interquartet hydrogen bonding, the cation-free GQ network was formed at RT, as is evident from Fig. 2a-c, which shows the STM image of cation-free GQ network with the superimposed DFT-optimized structural model.

**Effect of Graphene Nanoribbon Pre-Synthesis**

It has been known that due to reconstructed patterns and low-coordinated Au atoms, step edges represent preferential nucleation sites for the growth of supramolecular structures.[32,33,36,37] Therefore, guanine molecules have no choice other than making hydrogen bonds with a chain array of initial molecules at step edges that are not compatible with the GQ network's geometry. Based on our experiments, the disordered domains initiated from step edges can interfere with the GQ network's growth on the Mo-doped Au(111) surface (Fig. 2a). This interference can be spatially terminated by placing GNRs since a guanine molecule's interaction with GNR is weak. The effect of graphene nanoribbon pre-synthesis on guanine self-assembly was initially investigated on pristine Au(111) surface. The DFT calculation shows that the adsorption energy of a guanine molecule on 7-AGNR adsorbed on Au(111) is -0.94 eV, much weaker than the adsorption energy on Au(111) (-1.38 eV). Therefore, guanine molecules are hardly stacked on GNRs,[38,39] as evidenced in our experiment (Fig. 3b); hence 7-AGNR acts as neutral fences of cage area to grow guanine self-assembly.[25] The successful growth of GNRs on pristine Au(111) surface is confirmed by STM, as shown in Fig. 3a and S8. The guanine molecules were assembled within a region enclosed by GNRs rather than diffusing towards GNRs (Fig. 3b). It is consistent with that the binding energy of the free-standing GNR-guanine side-to-side configuration is -0.32 eV, while that of a guanine dimer is -0.65 eV. The STM image in Fig. 3c shows that guanine self-assembly is less affected by step edges, which are preferentially covered by GNRs. As the Au surface was free of Mo, we observed the disordered phase of G/9H for RT deposition (Fig. 3b) and pure G/7H close-packed structure for 400 K annealing (Fig. 3d). Guanine molecules were deposited at RT after the GNR pre-synthesis and Mo-doping to investigate both effects together. Extensive pure GQ islands on Mo-doped Au(111) surface with 7-AGNR pre-synthesis are shown in Fig. 4a. Decorating the surface with GNRs facilitated the formation of more extensive and homogeneous GQ networks by producing confined areas in which molecular networks nucleated without interacting with the GNR boundaries (Fig. 4a-d and Fig. S1d-j). GQ domains were surrounded inevitably by a minor number of disordered domains (Fig. 4b). GQ networks were stable at RT after eight weeks in UHV condition (Fig. 4c). Guanine molecules were not influenced by surface step edges after Mo-doping (Fig. 4d), as diffusion passage was blocked by GNRs. Step edges could be easily passivated by the H atoms released from GNR precursor (DBBA) during the cyclodehydrogenation step of GNR synthesis at 670 K.[40,41] Since the Mo concentration of Au(111) surface layer was minute, it did not affect the on-surface synthesis of 7-AGNRs.

**Annealing GQ networks and disordered domains**

Although both Mo-doping and GNR pre-synthesis effects induce the one-step formation of extensive GQ networks at RT, a minor number of disordered domains were still observable (Fig. 4a-b and S1h-j). To investigate the origin of the disordered domains, we annealed guanine self-assembly on the Mo-doped Au surface without GNR pre-synthesis (Fig. 2a). Disordered domains of guanine on Mo-doped Au(111) transformed into the close-packed structure at 400 K, while some disordered phase remained (Fig. 5a). The rise of disordered domains originated from not only the interference of assemblies initiated from step edges but also the locally overdoped Au(111) surface by Mo (Fig. 5a and S7a). This made adsorbed guanine molecules rather fixed with a lack of conformality to GQ lattice due to relatively strong covalent bonds with the substrate. The disordered domains on the Mo-doped Au surface finally transformed into the G/7H structure after annealing at 460 K (Fig. 5b). In contrast to the GQ network, the close-packed lattice had the flexibility to adapt with guanine molecules bonded at step edges or Mo atoms (Fig. S7b-c). DFT calculation also showed that the G/7H molecule makes a covalent bond with an embedded Mo atom in the Au surface layer in a face-down geometry with a vertical distance of 2.1 Å between O and Mo atom and -2.37 eV adsorption energy (Fig. 5c). For Mo-doped Au surface with GNR pre-synthesis, annealing at 460 K resulted in pure GQ networks and a few close-packed islands along with 7-AGNRs (Fig. 5d). At 500 K annealing, GNRs moved towards each other and did not mix/bond with GQ networks or close-packed G/7H structures. Small GQ networks (less than 10×10 nm$^2$ area) were no longer observable at 500 K annealing (Fig. 5e), either merged or gradually converted to the close-packed lattices. The higher annealing temperature of 520 K resulted in a complete conversion from the empty G-quartet to the G/7H structure with separated and unchanged GNRs (Fig. 5f and S7e).

**Electronic structure**

The role of Mo atoms in stabilizing the initial seed was further investigated from the electronic structure by STS measurement in collaboration with DFT calculations. To find the anchored molecule on the surface, we annealed the GQ structure on Mo-doped Au(111) surface (Fig. 2a) at 400 K. As a result, disordered domains converted to the close-packed networks (Fig. 5a). However, some disordered domains remained due to guanine molecules' covalent bonding with the surface defects. We measured the STS of G/9H molecules in the GQ network, G/7H molecules in the close-packed network, and guanine molecules in disordered domains (Fig. 6a). Although the LUMO peaks of the molecule in GQ and close-packed networks are out of the achievable range by STS, the recorded spectrum reveals a new, broad spectroscopic feature in the positive potential for the guanine molecule anchored on the surface. To unravel the origin of this broad spectroscopic peak, the projected density of states (PDOS) of a G/9H molecule in a GQ network and a G/7H molecule which have van der Waals interaction with Au(111) surface were calculated along with the PDOS for a covalently bonded G/9H molecule (in GQ lattice) with Mo-doped Au(111) surface (Fig. 6b). The newly emerged broad LUMO peak resulted from the covalent bonding between guanine molecule and surface (red curves in Fig. 6b). The additional coupling of $p_z$ orbitals of O atom of the anchored guanine molecule and the embedded Mo atom at Au surface (Fig. 6c) lowered the molecule's energy levels (Fig. 6b). Accordingly, LUMO's peak of the chemically bonded molecule with the surface became achievable by the STS and identified the anchored guanine molecule on the surface (red curve in Fig. 6a).

## Conclusions

In conclusion, we have successfully prepared the cation-free G-quartet network on Au(111) surface at RT by strengthening the first molecule-surface interaction and controlling the initial seed's stability and geometry. The pre-synthesis of GNRs on Mo-doped Au surface modified the surface domain spatially and chemically, which prompted effective nucleation and growth of GQ networks. The nucleation and growth mechanisms are unraveled from the interplay of STM imaging, STS measurements, and DFT calculations. GQs and GNRs relocated while stayed segregated from each other at various temperatures. The cation-free GQ network exhibited robust thermal stability against annealing up to 500 K. Interconversion of all networks to a more stable close-packed G/7H network in the presence of GNRs was achieved at 520 K. Our findings could open a new window for applications of graphene nanostructures and bimetallic surfaces in the stabilization of new emerging biomolecular structures, which are often being used as therapeutic targets in human diseases, including cancer therapy, and from a different point of view, for a new generation of DNA-based nanoelectronic circuits and miniaturized memory storages.

## Materials and methods
### Sample preparation

On-surface 7-AGNR synthesis on Mo-doped Au(111) surface was done using 10,10′-dibromo-9,9′-bianthracene (DBBA) precursor monomers (purchased from Sigma Aldrich, purity >98%) deposited via DCT at RT (Fig. S8).[42] The sample subsequently annealed gradually for dehalogenation below 370 K as well as polymerization at 470 K and cyclodehydrogenation at 670 K for 10 min,[40,41,43] in the same way as the GNR synthesis on bare Au(111) surface. The guanine 9H molecules (purchased from Sigma Aldrich, purity >98%) were first degassed thoroughly, then thermally sublimated onto the clean Au(111) substrate from a tantalum crucible heated to 500 K, resulting in a deposition rate of 0.01 molecule/min (0.01 ML/min) on bare Au(111), and 0.03 molecule/min (0.03 ML/min) on Mo-doped Au(111) surface with/without GNR pre-synthesis. For surface defect preparation, a molybdenum rod (from Omicron purity >99%) was installed on the sample holder to use as a sputtering target. Mo clusters deposited on the Au surface at 520 K (operating pressure $5.0 \times 10^{-6}$ mbar) were alloyed with Au surface during post-annealing first at 830 K/10 min for herringbone reconstruction of Au(111) surface, then at 700 K for 30 min to get large clean terraces and to remove any trace of Mo adatoms that might remain on Au surface (Fig. S3). The sample was transferred to the STM chamber subsequently within UHV to do STM measurements at 77K. WSxM and Vernissage software were used to process STM images.[44]

### SPM Characterization and Measurements

All STM experiments were performed in a UHV chamber (base pressure $1.0 \times 10^{-10}$ mbar) equipped with a low-temperature STM from Omicron Nanotechnology using electrochemically etched W tips. For all the experiments, the 200 nm thin film of Au(111) epitaxially grown on mica (Phasis, Switzerland) was used as the substrate. Standard Ar$^+$ sputtering/annealing cycles were applied to prepare an atomically clean surface. d$I$/d$V$ measurements were recorded using a lock-in amplifier with a modulation frequency of 701 Hz and a bias modulation amplitude of $V_{ac}$ = 24 mV. d$I$/d$V$ point spectra were recorded under open feedback loop conditions.

### Surface Characterization and Measurements

TEM sample preparation was performed using the FIB-SEM technique. The gallium ion beam was employed to cut a thin slab of bulk Au. HR-TEM/HAADF STEM verified pure Mo doped at the Au surface, and EDS elemental mapping analysis showed Mo concentration in the subsurface layer. TOF-SIMS technique is used to analyse the composition of the Mo-doped Au surface alloy (Fig. S2g).

### Density functional theory (DFT) calculations

DFT calculations were performed using the Vienna Ab initio Simulation Package (VASP). Tkatchenko and Scheffler (TS) dispersion correction[45] with the Perdew-Burke-Ernzerhof (PBE) exchange functional[46] were used to describe the adsorption of organic molecules on the Au(111) surface. In Au(111) slab supercell, four layers of Au atoms were considered, and the top two layers were optimized. More than 12 Å of vacuum space were considered to avoid the interaction between periodic slabs. The plane-wave energy cut-off was 550 eV.

## Conflicts of interest

There are no conflicts to declare.

## Acknowledgements

This work was supported by the National Research Foundation (NRF) of Korea (National Honor Scientist Program: 2010-0020414, Basic Science Research Program: 2018R1D1A1B07045983), and KISTI (KSC-2019-CRE-0253, KSC-2020-CRE-0049, KSC-2020-CRE-0146). A.R.G would like to thank Saeed Pourasad for a fruitful discussion on theoretical simulations.

**Table 1. GQ network formation energies:** free-standing, on pristine Au(111), and on Mo-doped Au(111) surface with one, two, or three Mo atoms embedded at the surface layer. Free standing lattice formation energy per molecule: $E_f = [E_{2D\ lattice} − 4×E_{molecule}]/4$; lattice formation energy on the surface per molecule: $E_f = [E_{2D\ lattice@surface} − 4×E_{molecule} − E_{surface}]/4$.

| GQ network | Formation energy (eV/molecule) |
|---|---|
| Free-standing | -1.27 |
| On bare Au(111) surface | -1.82 |
| On Mo-Au(111) surface (single Mo atom at surface) | -1.96 |
| On Mo-Au(111) surface (two Mo atoms at surface) | -2.17 |
| On Mo-Au(111) surface (three Mo atoms at surface) | -2.12 |

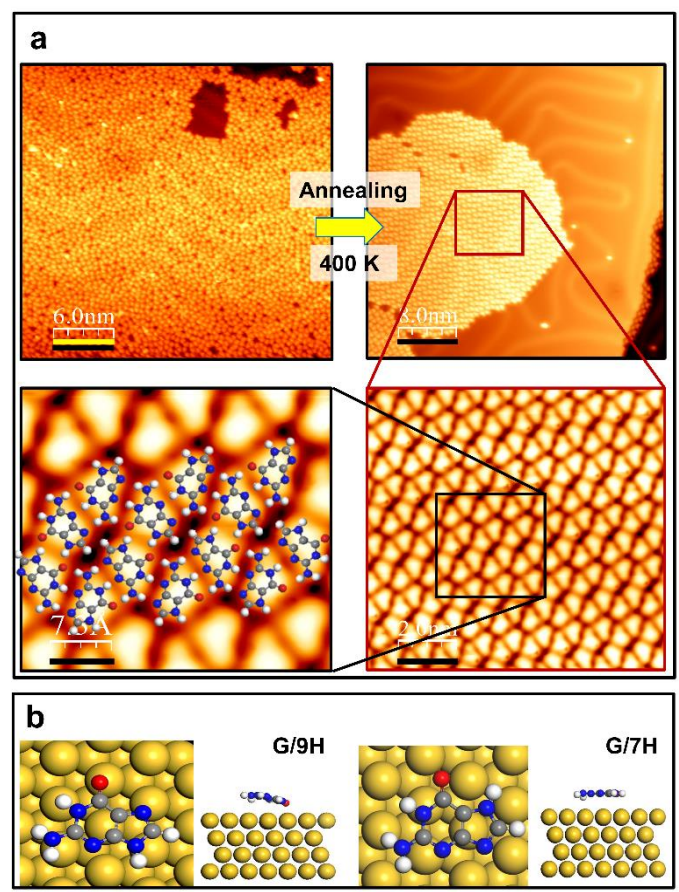

**Fig. 1 Guanine on pristine Au(111) surface.** (a) STM images before and after annealing at 400 K (upper panel). The latter is shown in smaller scales (2.0 nm and 7.3 Å) in the lower panel, depicting that the disordered phase of self-assembled guanine is transformed to the close-packed network on Au(111) with the DFT-optimized structural model superimposed. Scanning conditions (before annealing): tunneling current $I_t = 0.055$ nA, sample voltage $V_s = +0.9$ V; Scanning conditions (after annealing): $I_t = 0.15$ nA, $V_s = +0.9$ V. (b) Top and side views of DFT-predicted geometries for the single G/9H (left) and G/7H (right) molecule adsorption on Au(111). Single-molecule adsorption energy ($E_a = E_{molecule@surface} − E_{surface} − E_{molecule}$) on the surface is calculated to be -1.38 eV (shortest vertical distance $h = 2.6$ Å) for G/9H molecule, while -1.24 eV ($h = 3.2$ Å) for G/7H molecule). H: white, C: gray, N: blue, O: red, Au: yellow.

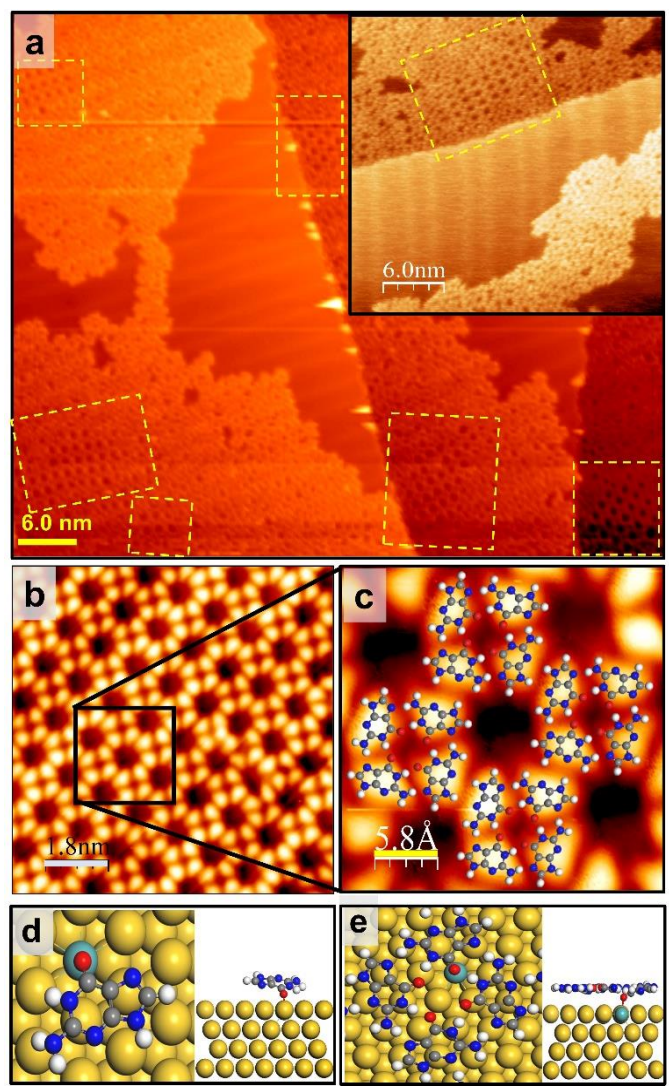

**Fig. 2 G/9H molecule adsorption and GQ network formation on Mo-doped Au(111).** (a) RT self-assembly of GQ networks coexisting with the disordered domains when Mo is embedded at the Au(111) surface. GQ networks on the Mo-doped Au surface are rare because there is a limited chance for them to grow without the interference of disordered domains coming from the step edges. The inset image shows a higher resolution of GQ network. GQ networks are highlighted in yellow dashed lines. Scanning conditions: $I_t = 0.029$ nA, $V_s = +0.9$ V. (b) Close-up high-resolution STM images of guanine self-assembly on Mo-doped Au(111). (c) The GQ network geometry obtained by DFT is superimposed. Scanning *conditions*: $I_t = 0.125$ nA, $V_s = +0.8$ V. (d-e) Top and side views of DFT-predicted geometry for (d) the single G/9H molecule adsorption ($E_a = -2.42$ eV and $h = 2.0$ Å) and (e) the GQ network formation energy $E_f = -1.96$ eV/molecule on Mo-doped Au(111). (also see Table 1 and Fig. S4) H: white, C:

gray, N: blue, O: red, Mo: azure, Au: yellow. Scale bar: 6.0 nm in panel a, 1.8 nm in panel b, 5.8 Å in panel c.

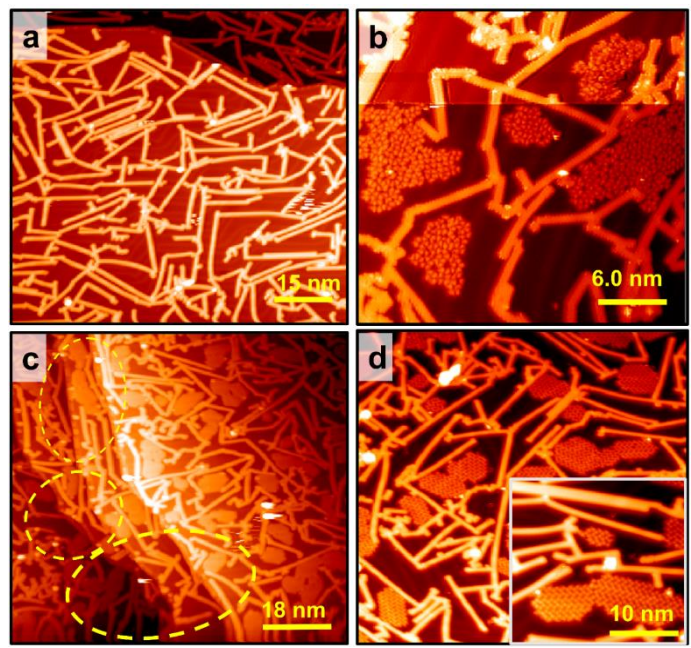

**Fig. 3 Effect of GNR pre-synthesis on guanine nucleation and growth on pristine Au(111) surface at RT.** (a) STM image of 7-AGNRs synthesized on Au(111). Scanning conditions: It = 0.03 nA, Vs = +0.75 V. (b) Guanine monolayer islands (disordered phase) grown on Au(111) until they reached GNR edges. Self-assembled guanine nucleated on the pristine Au surface between GNRs without binding them. Guanine molecules were not stacked on the GNR surface. Scanning conditions: It = 0.015 nA, Vs = +1.10 V. (c) Step edges are not preferential nucleation zones after GNR pre-synthesis, as shown with yellow dashed lines. Scanning conditions: It = 0.003 nA, Vs = +0.8 V. (d) G/7H close-packed structure after annealing the disordered phase at 400 K. The inset image shows a higher resolution of the zigzag structure of G/7H tautomers. Curing the disordered phase on pristine Au surface by GNR pre-synthesis is not sufficient to nucleate a GQ network's seed at RT, even though the destructive effect of Au step edges is eliminated. Scanning conditions: It = 0.02 nA, Vs = +0.9 V.

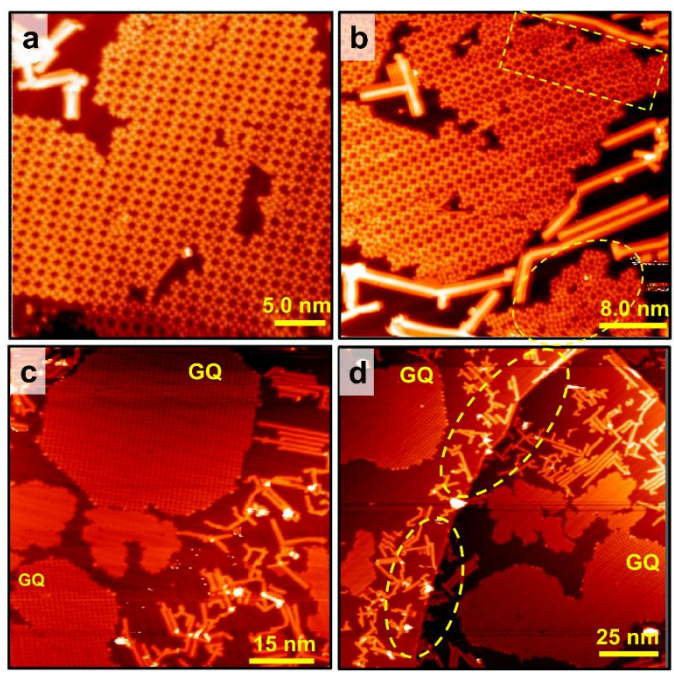

**Fig. 4 Effect of GNR pre-synthesis on GQ network's growth on Mo-doped Au(111) surface at RT.** (a) Extensive G-quartet networks coexisted disordered domains at network edges formed on Mo-doped Au surface at RT with 7-AGNR pre-synthesis. Low-concentration Mo doped at Au surface along with GNR pre-synthesis induced RT growth of GQ networks all over the surface. Scanning conditions: It = 0.023 nA, Vs = +0.9 V. (b) Extensive GQ networks are enclosed by 7-AGNRs and separated from disordered domains. Disordered domains terminated by GNRs are shown with yellow dashed lines. GQ networks are encircled by 7-AGNRs neutral fences. Scanning conditions: It = 0.03 nA, Vs = +0.9 V. (c) Extensive GQ networks on Mo-doped Au surface with 7-AGNRs were stable after eight weeks in UHV condition at RT, whereas disordered domains were not stable at RT and converted to the close-packed structure after this time. (d) Step edges are shown with yellow dashed lines, which are not preferential nucleation zones for guanine molecules after GNR pre-synthesis on Mo-doped Au(111) surface. Scanning conditions: It = 0.03 nA, Vs = +0.8 V. Scale bar: 5.0 nm in panel a, 8.0 nm in b, 15.0 nm in panel c, and 25 nm in panel d.

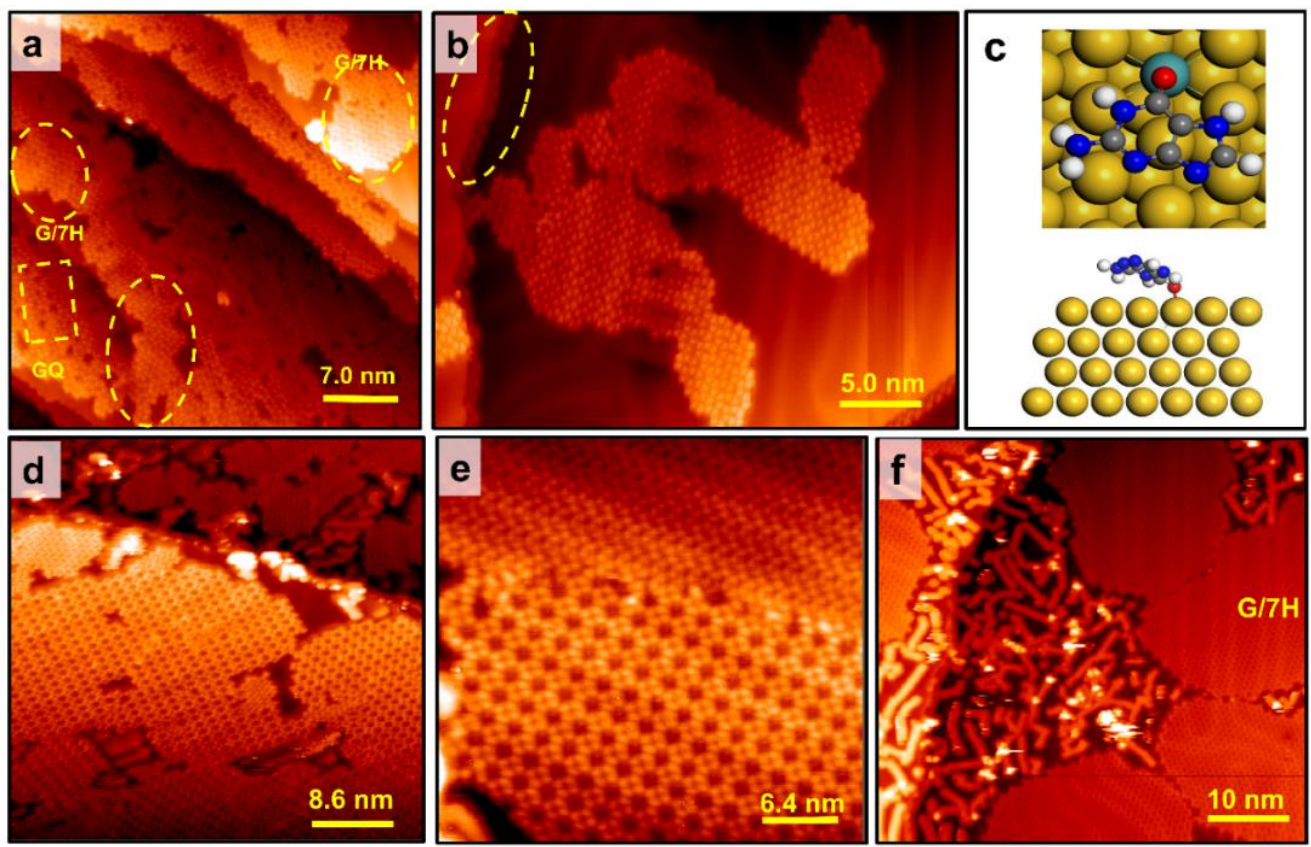

**Fig. 5 Annealing the guanine self-assemblies on Mo-doped Au surface.** (a) At 400 K, a disordered domain was partially transformed into the close-packed structure, while some disordered phase remains due to relatively strong covalent bonds with Mo embedded at Au surface (on Mo-doped Au surface without GNR pre-synthesis). GQ networks remained at this temperature. Ordered domains (GQ network and close-packed islands) are highlighted in yellow dashed lines. Scanning conditions: $I_t$ = 0.019 nA, $V_s$ = +0.9 V. (b) Guanine self-assemblies on Mo-doped Au were fully transformed into G/7H structure, which has the flexibility to adapt with guanine molecules bonded at step edges or Mo atoms. Scanning conditions: $I_t$ = 0.02 nA, $V_s$ = +0.9 V. (c) Top and side views of DFT-predicted geometry for single G/7H molecule adsorption ($E_a$ = -2.37 eV and $h$ = 2.1 Å) on Mo-doped Au(111) surface. H: white, C: gray, N: blue, O: red, Mo: azure, Au: yellow. (d) At 460 K, disordered domains on Mo-doped Au with GNR pre-synthesis were fully transformed into G/7H structure, while pre-formed GQ networks were stable at this temperature. Scanning conditions: $I_t$ = 0.02 nA, $V_s$ = +1.0 V. (e) At 500 K, the G/7H lattice coexists with GQ network, while small islands of GQ domains no longer exist. Scanning conditions: $I_t$ = 0.044 nA, $V_s$ = +0.9 V. (f) At 520 K, large islands of the close-packed structure are formed. GNRs were relocated while stayed segregated from guanine assemblies and gathered around step edges or between large islands of G/7H networks. Scanning conditions: $I_t$ = 0.035 nA, $V_s$ = +0.9 V. Scale bar: 7.0 nm in panel a, 5.0 nm in panel b, 8.6 nm in panel d, 6.4 nm in panel e, and 10 nm in panel d.

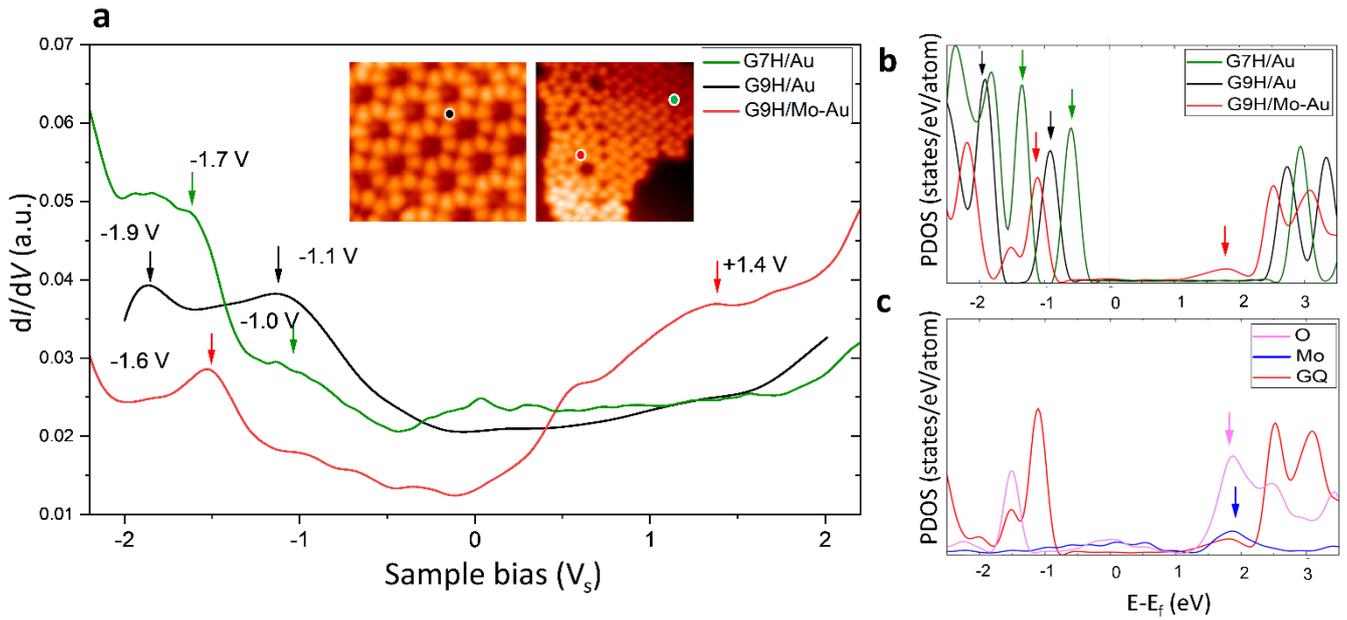

**Fig. 6 Electronic structure of guanine molecules on Au(111) surface.** (a) The black (green) curve shows d$I$/d$V$ point spectroscopy data collected from the G/9H (G/7H) molecule in a GQ (close-packed) network on Mo-doped Au(111). The red curve is d$I$/d$V$ point spectroscopy data collected from a G/9H molecule in a disordered domain on Mo-doped Au surface (Inset image shows a GQ network (left), and a zigzag lattice and anchored guanine molecules (right) derived from a disordered domain after annealed at 400 K; also see Fig. 5a). The spectroscopy parameters are $V_{ac}$ = 24 mV and $f$ = 701 Hz, d$I$/d$V$ point spectra were recorded under open feedback loop conditions. Scanning conditions: $I_t$ = 0.155 nA, $V_s$ = +1.0 V. All STS data shown here were obtained at T = 77 K. (b) DFT projected DOS (PDOS) for a G/9H molecule in a GQ network on pristine Au(111) (black), Mo-doped Au(111) (red) surfaces (Fig. 2e), and a G/7H molecule (green) on pristine Au(111) surface (Fig. 1b). (c) The guanine molecule's chemical bond with the surface results in additional p$_z$ orbital coupling of Mo and O atoms (mentioned by blue and purple arrows) and yields a newly emerged broad LUMO peak (mentioned by a red arrow in b). This peak can be defined as a characteristic for the guanine molecule's covalent bonding with the Au surface. Blue and purple arrows in (c) show the origin of the LUMO peak mentioned by a red arrow in (b). Green, black, and red arrows in (a) and (b) are peer-to-peer related.